\documentstyle[aps]{revtex}
\begin{document}
\draft

\title{Tidal Stabilization of Rigidly Rotating, Fully Relativistic Neutron
Stars}
\author{Kip S. Thorne}
\address{Theoretical Astrophysics, California Institute of Technology,
Pasadena, CA 91125}
\date{Submitted 17 June 1997; Revised 10 January 1998}
\maketitle
\begin{abstract}
It is shown analytically that an external tidal gravitational field 
increases the secular stability of
a fully general relativistic, rigidly rotating neutron star that is near
marginal stability, protecting it 
against gravitational collapse.  This stabilization is shown to result from the
simple fact that the energy $\delta M({\cal Q},R)$ required to raise a tide on
such a star, divided by the square of the tide's quadrupole moment $\cal Q$, is
a decreasing function of the star's radius $R$, $(d/dR)[\delta M({\cal
Q},R)/{\cal Q}^2] < 0$ (where, as $R$ changes, the star's structure is changed
in accord with the star's fundamental mode of radial oscillation). If 
$(d/dR)[\delta M({\cal Q},R)/{\cal Q}^2]$
were positive, the tidal coupling would destabilize the star.
As an application, a rigidly rotating, 
marginally secularly stable neutron star in an inspiraling binary system 
will be protected against secular collapse, and against dynamical collapse,
by tidal interaction 
with its companion.  The ``local-asymptotic-rest-frame'' tools used 
in the analysis are somewhat unusual and may be powerful in other studies of 
neutron stars and black holes interacting with an external environment.
As a byproduct of the analysis, in an appendix the influence of tidal 
interactions on mass-energy conservation is elucidated.
\end{abstract}

\narrowtext
\twocolumn

\section{Introduction and Summary}
\label{sec:intro}

Wilson, Mathews, and Maronetti \cite{WilsonMathews} have carried out 
fully relativistic numerical simulations of the radiation-reaction-induced
inspiral of a binary neutron star system.  To make their computations 
tractable, they employed several approximations of ill-understood accuracy.  
The stars in their simulations were identical and were near the maximum 
allowed mass for an isolated neutron star.  Correspondingly, when the stars
were far apart in their orbit, each was stable against gravitational 
collapse.  Surprisingly, as the stars spiraled inward, the simulations
indicated that their gravitational interaction destabilized them, triggering
them to collapse before
their inspiral ended.  The magnitude of the destabilization and 
mathematical arguments to explain it
\cite{WilsonMathews} suggest that it should show up in
the first post-Newtonian approximation to general relativity. 

Several researchers have argued that this surprising destabilization is wrong:
Lai \cite{Lai} has shown that tidal
interactions between two nearly Newtonian stars will tend to stabilize 
them against gravitational
collapse, not destabilize them.  Lai's stabilization effect is formally of
Newtonian origin, but because of the compactness of neutron stars, its 
magnitude is of much higher post-Newtonian order.  Wiseman \cite{Wiseman} 
has elucidated Lai's conclusion by showing that at first post-Newtonian 
magnitude, the stars' gravitational interactions do not alter their individual 
central densities;
and Brady and Hughes \cite{BradyHughes} have shown the same at first order
in the mass ratio $M_2/M_1$ when the two stars are fully relativistic
and one is much less massive than the other, $M_2 \ll M_1$.  Baumgarte et.\
al.\ \cite{Baumgarte} have carried out numerical simulations of the fully
relativistic equilibrium states of a binary neutron star system in synchronous,
circular orbit --- simulations analogous to those of Wilson, Mathews, and 
Maronetti. Not only do these simulations show no sign of interaction-induced
collapse; when combined with ``turning-point'' criteria for secular 
stability, they actually reveal a stabilization of the stars.  

On the basis of
these analyses, it seems likely that the destabilization seen by
Wilson, Mathews and Maronetti does not occur in reality, but instead is an 
artifact of poor-accuracy approximations or is due to some error in their 
computations. 

In this paper we present another analysis that reveals stabilization,
not destabilization.  Our justification for yet another paper on
this subject is two-fold:  
{\sl First}, our analysis has broader validity than previous
ones---it is fully relativistic, not post-Newtonian; 
and unlike the two previous fully relativistic analyses of
stability \cite{WilsonMathews,Baumgarte}, it is fully analytic
and not based on numerical simulations, it permits the stars to rotate with
arbitrary angular velocity (though with spins aligned with the orbital angular
momentum), and it allows an arbitrary mass ratio. 
{\sl Second}, our analysis employs an unusual approach, which may be 
useful for other problems in fully relativistic binary evolution:
it is formulated in the local asymptotic rest frame of one of the two stars and
employs energy and angular momentum arguments that relate to that star alone
and not to the binary system as a whole.  Although this approach is unusual 
within general relativity, it is well known in Newtonian and post-Newtonian
theory.  It, in fact, is a relativistic generalization of Lai's \cite{Lai}
post-Newtonian proof of stabilization. 

This paper is organized as follows:  In Sec.\ II we treat an idealized problem
that illustrates our method: the stabilization of a non-spinning neutron
star placed in a non-rotating external tidal gravitational field.  Then in 
Sec.\ III we generalize to a spinning star and rotating tidal field, and as an
application we deduce the {\sl secular} stabilization of a spinning
neutron star in an inspiraling binary.  In Sec.\ IV we argue from this secular
stabilization result that, if an inspiraling binary's neutron stars are 
secularly stable at large orbital radii, then they cannot be {\sl dynamically}
destabilized during the inspiral; and we make some concluding remarks.
In an appendix we elucidate the influence of tidal interactions on energy
conservation, in the Newtonian limit.  The relativistic version of this issue
is a central aspect of the proof of stability given in the
body of the paper.

Throughout we use geometrized units in which the speed of light and Newton's
gravitation constant are unity. 

\section{Static Star and Static Tidal Field}
\label{sec:Static}

\subsection{Momentarily Static, Spherical Star}
\label{sec:StaticSpherical}

\begin{figure}
\vspace{10.4pc}
\special{hscale=70 vscale=70 hoffset=30 voffset=5
psfile=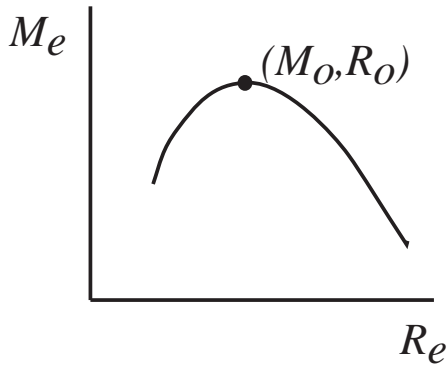}
\caption{The Mass-Radius curve for static, spherical equilibrium 
configurations (neutron stars) with some equation of state $P(\rho)$.} 
\label{fig:MeRe}
\end{figure}

Consider a family of nonrotating, spherical neutron stars characterized by 
a one-parameter equation of state $P = P(\rho)$, where $P$ is pressure and
$\rho$ is density of total mass-energy.  We shall refer to these stars as {\sl
equilibrium configurations} and shall denote their masses (as measured by the
Keplerian orbits of distant planets) by $M_e$, their total number of baryons
by $N_e$, and their radii (defined as circumference$/2\pi$) by $R_e$.  The
mass-radius relation $M_e(R_e)$ of these equilibrium configurations has the
qualitative form shown in Fig.\ \ref{fig:MeRe}, and $N_e(R_e)$ has a similar
shape; cf.\ Refs.\ \cite{HTWW,ShapiroTeukolsky}.

As is well known, the equilibrium configuration of maximum mass (solid dot in
Fig.\ \ref{fig:MeRe}; ``{\sl critical configuration}''; 
mass $M_o$ and radius $R_o$) 
is secularly marginally 
stable:  It possesses a zero-frequency mode of expansion (or contraction) 
that takes it horizontally in
the figure to another equilibrium configuration with the same mass.  Equilibria
on the larger-radius side of the critical configuration are secularly
stable; those on the lower-radius side are secularly unstable 
\cite{HTWW,ShapiroTeukolsky}. 

\newpage
\null

By {\sl secular} is meant a mode of stellar deformation which is slow enough
for pycnonuclear reactions
(pressure-induced nuclear reactions) to keep its
matter always at the endpoint of nuclear evolution, so the pressure and density
changes experienced by the stellar matter follow the same equation of state
$P(\rho)$ as characterizes the equilibrium configuration itself.  
The slowest of the pycnonuclear reactions are ``modified URCA reactions''
(essentially $\beta$ and inverse-$\beta$
decays), driven as the star deforms by the rising or falling Fermi energies of
the star's electrons, protons, and neutrons; they can require timescales of 
minutes or longer to equilibrate near and below nuclear densities, so in
principle the 
stellar deformations can be secular only on timescales longer than this. 
(In practice, these slow reactions have only a weak effect on the equation of
state near and above nuclear densities, so their slowness is often ignored for
near critical neutron stars.)

Faster ({\sl dynamical}) motions, in which some of the pycnonuclear reactions 
do not go to 
completion, will be characterized by a stiffer equation of state 
(higher adiabatic index) than secular motions and thus will be more stable. 
Correspondingly, all equilibria on the large-$R_e$ branch, being
secularly stable, must also be dynamically stable, as must be the critical
configuration itself.  For an ancient, further discussion of these
issues but in different language, see Ref.\ \cite{MeltzerThorne}

In the next subsection we shall study the influence of an external tidal field
on the secular stability of configurations that are nearly critical (nearly at
the maximum of the mass-radius curve).  As an aid in that study, it will be
useful to consider momentarily static, spherical stars that are deformed 
slightly away from equilibrium.  For a star containing $N$ baryons (with $N$
very nearly equal to $N_o$), we obtain such a non-equilibrium
configuration as follows:  Begin with the equilibrium configuration that has
$N_e = N$ and radius $R_e(N_e)$.  Expand it or contract it slightly to the
desired new radius $R$.  In this deformation, displace each fluid element by an
amount proportional to the equilibrium configuration's fundamental-mode
eigenfunction $\vec \xi (\vec x)$ of secular vibration, and then remove all 
kinetic energy of
deformation.  The resulting deformed configuration will have a mass $M(N,R)$
that differs slightly from the equilibrium mass $M_e(N_e=N)$. 

In practice, the mass of the deformed star can be computed using not the 
star's true eigenfunction, but rather the eigenfunction $\vec \xi_o (\vec x)$ 
of the
zero-frequency mode of the critical configuration.  This is because the mass is
an extremum with respect to deformations of the star \cite{HTWW}; the two 
eigenfunctions $\vec \xi$ and
$\vec \xi_o$ differ by fractional amounts of order $n \equiv (N-N_o)/N_o$, so
the masses of the configurations obtained by deformations to radius $R$ via the
$\vec \xi$ motion and the $\vec \xi_o$ motion will differ by a fractional
amount of order $n^2$ -- which is never of interest in this paper.

The configurations obtained by the above construction are characterized by two
parameters, $(N,R)$; and their masses are functions of these parameters,
$M(N,R)$.  

For ease of analysis, we shall now convert to dimensionless variables:
\begin{equation}
m\equiv {{M-M_o}\over M_o}\;, \quad n\equiv {{N-N_o}\over N_o} \;, 
\quad r\equiv {{R-R_o}\over R_o}\;.
\label{eq:dimensionless}
\end{equation}
These variables characterize a configuration's fractional deviations from the
critical configuration.

\begin{figure}
\vspace{11pc}
\special{hscale=70 vscale=79 hoffset=40 voffset=5
psfile=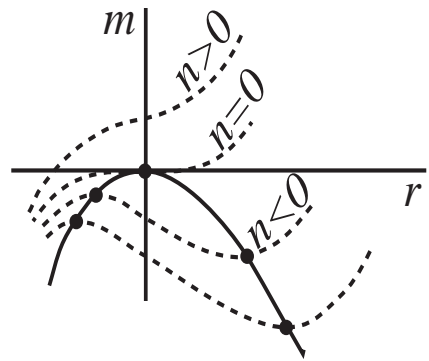}
\caption{Solid curve: the dimensionless mass-radius relation for the
equilibrium configurations of Fig.\ \protect\ref{fig:MeRe}.  Dashed curves: 
the dimensionless mass-radius relations $m(n,r)$ for configurations of
fixed baryon number $n$ that are obtained from an equilibrium configuration via
deformation along the fundamental eigenfunction of radial secular motion.}
\label{fig:mnr}
\end{figure}

Figure \ref{fig:mnr} shows the dimensionless
mass-radius relation $m(n,r)$ for configurations
with fixed baryon number $n$ (dashed curves), along with the equilibrium
configurations (solid curve).  Because the equilibria on the positive-$r$
branch are stable against secular deformations (with eigenfunction 
$\vec \xi$), they lie at minima of the dashed curves; because those on the
negative-$r$ branch are unstable, they lie at maxima of the dashed curves.  

These dashed curves, when expressed as a power series, have the following 
form:
\begin{equation}
m = ar^3 + (b_0 + b_1 r)n\;,
\label{eq:mnr}
\end{equation}
where higher-order terms are of no importance in this paper, and where $a$,
$b_0$, and $b_1$ are all positive.  This form is
dictated by the demand that for $n=0$ the dashed curve be flat at $r=0$, and
that for $n<0$ the positive-$r$ branch have a minimum (stable equilibrium)
and the negative-$r$
branch have a maximum (unstable), and that for $n>0$ there be no equilibria 
at all (no extrema of the dashed curves).  

The equilibrium configurations are located at
the extrema of these mass-radius curves, i.e.\ at locations where $(\partial m
/ \partial r)_n = 0$, which yields the following equilibrium 
relations:
\begin{eqnarray}
n_e &=& {-3a\over b_1}r_e^2\;, 
\quad m_e = {-3ab_0\over b_1}r_e^2 -2ar_e^3\;, \nonumber \\
\quad m_e &=& b_0n_e \mp 2 {(b_1/3)^{3/2} \over a^{1/2}} (- n_e)^{3/2}\;,
\label{eq:equi}
\end{eqnarray}
where the upper sign is for the right branch (stable stars; lower mass at fixed
$n_e$) and the lower sign, for the left branch (unstable stars; higher mass at
fixed $n_e$). 

The coefficients $a$, $b_0$, and $b_1$ are determined as follows in terms of
the equilibrium configurations:  As is well known---cf.\ Eq.\ (28) of Ref.\
\cite{HTWW}---the mass-energy required to create one baryon and inject it into 
an arbitrary location in an equilibrium configuration, in local 
thermodynamic equilibrium with the matter there, is $dM_e/dN_e = 
\mu_B \sqrt{1-2M_e/R_e}$,
where $\mu_B$ is the rest mass of one baryon at the star's surface
(1/56 the mass of an $^{56}$Fe nucleus if the star's matter has been
``catalyzed to the endpoint of thermonuclear evolution'' \cite{HTWW}).  
Evaluating this ``injection energy'' for the critical configuration, switching
to dimensionless variables, and comparing with $(dm_e/dn_e)_o = b_0$  
[Eq.\ (\ref{eq:equi})], we see that
\begin{mathletters}
\label{eq:coefficients}
\begin{equation}
b_0 = (\mu_B N_o / M_o) \sqrt{1-2M_o/R_o} \sim 0.8\;.
\label{eq:b0}
\end{equation}  
Performing the same calculation slightly away from the critical configuration,
we obtain
\begin{equation}
b_1 = {\mu_B N_0 /R_0 \over \sqrt{1-2M_o/R_o}} \sim 0.6 \;.
\label{eq:b1}
\end{equation}
The remaining coefficient, $a$, is determined by the curvature $(d^2M_e /
d{R_e}^2)_o$ of the
equilibrium mass-radius relation at its critical point; cf.\ Eq.\ 
(\ref{eq:equi}):
\begin{equation}
a = {R_o \over 6(1-2M_o/R_o)} \left( -d^2M_e \over dR_e^2 \right)_o \sim 1\;.
\label{eq:a}
\end{equation}
\end{mathletters}
In Eqs.\ (\ref{eq:coefficients}) and later equations in this paper,
the numerical values have been inferred, with uncertainties typically
no worse than
a factor two, 
from the
equilibrium configurations for plausible equations of state
\cite{ShapiroTeukolsky}.

\subsection{Static, Tidally Deformed Star}
\label{sec:StaticTidal}

We now place a near-critical neutron star in a static, external
tidal gravitational field, which we characterize by the space-time-space-time
components of its Riemann tensor, $R_{j0k0} \equiv {\cal E}_{jk}$ \cite{rmp}.  
This tidal field will deform the star, i.e.\
will gravitationally ``polarize'' it, giving it a gravitational mass quadrupole
moment ${\cal I}_{jk}$.  The tidal field ${\cal E}_{jk}$, quadrupole moment 
${\cal I}_{jk}$, and total stellar mass $\cal M$ (including the deformation
energy and the energy of interaction between the deformation and the tidal
field) all show up as coefficients in a power
series expansion of the spacetime metric in the star's {\sl local asymptotic
rest frame} \cite{ThorneHartle}.\footnote{Reference \cite{ThorneHartle} is the
principal conceptual and mathematical foundation for this paper's analysis. 
The physical concepts that underlie Ref.\ \cite{ThorneHartle} and this
paper, including the validity of the
equivalence principle for ``extended''
self-gravitating bodies such as neutron stars, 
date back to John Wheeler's discussion of equations of motion in
Sec.\ 20.6 of MTW\cite{MTW} and to references therein.}
For example, in harmonic (deDonder)
coordinates that are attached to the star's center of mass, 
the time-time metric component 
has the form \cite{rmp,ThorneHartle,Gursel}
\begin{eqnarray}
g_{00} &=& - 1 + 2{{\cal M}\over r} - 2 {{\cal M}^2\over r^2} 
+ 2 {{\cal M}^3\over r^3} 
+3 {{\cal I}_{jk} n^j n^k \over r^3} + \ldots \nonumber \\
&&- {\cal E}_{jk} n^j n^k r^2 + \ldots\;,
\label{eq:acmc}
\end{eqnarray}
where the first ``$\ldots$'' denotes higher-order terms in $1/r$ and the second
``$\ldots$'' denotes higher-order terms in $r$.  Here, and only here, $r$ is
coordinate radius computed as though the spatial coordinates $x^j$ were 
precisely Cartesian (elsewhere in this paper $r \equiv (R-R_o)/R_o$ is the 
star's dimensionless radius); 
and $n^j \equiv x^j/r$ is the ``unit radial vector''.  
The local asymptotic rest frame, where the expansion (\ref{eq:acmc}) is 
valid, is the region from the neutron star's surface out to a distance where
the external tidal field can no longer be regarded as uniform. 

For simplicity, and in accord with the case of a tidal 
field produced by a distant binary companion, we shall assume that the 
tidal field is axisymmetric; and we shall choose our (nearly Cartesian)
harmonic coordinates so its
symmetry axis is along the $x^3$ direction.  Then the 
induced quadrupole moment 
also possesses this symmetry, and the components of the tidal field and
quadrupole moment take the form
\begin{equation}
{\cal E}_{xx} = {\cal E}_{yy} = - {1\over2} {\cal E}_{zz} \equiv {\cal E},
\quad
{\cal I}_{xx} = {\cal I}_{yy} = - {1\over2} {\cal I}_{zz} \equiv -{\cal Q}.
\label{eq:EQ}
\end{equation}
Here the signs are chosen such that ${\cal E}$ and ${\cal Q}$ are both
positive.

In the next few paragraphs, culminating with Eq.\ (\ref{eq:Qe}), we shall
compute the magnitude $\cal Q$ of the quadrupole moment that is
induced by a given tidal field ${\cal E}$.  We do so by the following
thought experiment:  Begin with a near critical, spherical configuration of 
mass $M$, baryon number $N$, and radius $R$, and turn on the tidal field
$\cal E$
without letting the star deform.  Then allow the star to deform of its own
accord, in response to the fixed tidal field $\cal E$.  
The deformation will develop on 
the timescale of the star's $f$-mode quadrupolar oscillations,\footnote{The 
tidal field, having no radial nodes in its tidal force pattern,
will primarily excite the node-free $f$-mode; it will couple far more weakly to
the lower-frequency, longer-timescale $g$-modes 
\cite{Lai1,ReiseneggerGoldreich,Shibata}.}
$\sim 0.3$msec, which is
far faster than the star's radial motions; thus, its angle-averaged radius $R$
will remain essentially unchanged during the deformation.  
As the deformation proceeds, i.e.\ as $\cal
Q$ grows, the tidal field does work $W({\cal Q,E})$ on the star, i.e. it 
increases the star's total mass-energy (excluding quadrupole-tidal interaction
energy) by that amount.  Some portion $\delta
M({\cal Q},R)$ of the work $W$
goes into producing the stellar deformation (pushing mass up in
the polar regions and down in the equatorial regions); the rest goes into
kinetic energy of quadrupolar vibrations.  Suppose that we extract the kinetic
energy; then the star will settle down
into the deformed configuration that has released the most kinetic energy,
i.e.\ the configuration that minimizes the {\sl potential energy function}
\begin{equation}
V(N,R,{\cal Q,E}) \equiv M(N,R) + \delta M({\cal Q},R) - W({\cal Q,E})\;.
\label{eq:V}
\end{equation}

The quadrupolar deformation 
energy $\delta M({\cal Q},R)$ can be deduced in cgs units by dimensional
arguments.  It obviously must be quadratic in 
$\cal Q$, proportional to Newton's gravitation constant $G$ and independent of
the speed of light $c$, which means it must have the form $\delta M = \beta G
{\cal Q}^2 / R^5$ where $R$ is the configuration's radius (the only lengthscale
available other than those, of order $R$, that characterize the star's internal
structure).  Here $\beta$ is a dimensionless coefficient of order unity that
depends on dimensionless aspects of the star's internal
structure.  In geometrized units ($G=c=1$), this mass-energy of deformation is
\begin{equation}
\delta M({\cal Q},R) = {\beta \over R^5}{\cal Q}^2\;.
\label{eq:MQ2}
\end{equation}
This is the same expression as one obtains in Newtonian theory.  It can be
understood, in Newtonian language, as the gravitational energy 
$\sim (\Delta M)^2 /R$ of
an excess mass $\Delta M \sim {\cal Q}/R^2$ moved into the star's polar 
region a little higher than radius $R$,
and a corresponding mass deficit $-\Delta M$ in the star's equatorial
region a little lower than radius $R$.  As an example, for a uniform-density 
Newtonian star \cite{Kelvin}, $\beta = 3$. 

The work $W({\cal Q, \cal E})$, which the fixed tidal field $\cal E$ does on 
the 
star as its quadrupole moment grows from zero to the value $\cal Q$, can
be deduced by examining the flow of energy through the star's local
asymptotic rest frame.  That energy flow can be described by the 
Landau-Lifshitz pseudotensor \cite{LL} for the metric 
(\ref{eq:acmc}) with $\cal E$
fixed and $\cal Q$ time varying.  By integrating the pseudotensor
over a sphere
${\cal S}$ in the star's local asymptotic rest frame, 
we obtain the rate of change of the star's total
mass-energy [the quantity $\cal M$ appearing in the metric (\ref{eq:acmc})]: 
\begin{equation}
{d{\cal M}\over dt} = - \int_{\cal S}(-g) t_{LL}^{j0} n_j d^2A\;;
\label{eq:LLint}
\end{equation}
cf.\ Secs.\ 20.2 and 20.3 of Ref.\ \cite{MTW}.
This $d{\cal M}/dt$ 
consists of two parts: the rate $(d/dt) W({\cal Q}, {\cal E})$
that work is done by the external field on the star, and the rate of change
$dE_{\rm
int}/dt$ of the interaction energy 
of the external field and the stellar deformation
\begin{equation}
{d {\cal M}\over dt} = {d W({\cal Q}, {\cal E})\over dt} + 
{d E_{\rm int}\over dt}\;.
\label{eq:split}
\end{equation}
This split of $d {\cal M}/dt$ into two parts is elucidated 
in the Appendix and will be analyzed at greater length in a subsequent paper
\cite{PurdueThorne}. The integral (\ref{eq:LLint}) has been evaluated by
Zhang \cite{Zhang} using techniques described in Ref.\ \cite{ThorneHartle}
and foundations laid in \cite{Zhang1}.
The result, in a general situation where both $\cal E$ and $\cal Q$ may be
changing, is\footnote{Zhang does not give explicitly the numerical coefficient
$-1/10$, since the perfect time derivative term is not of interest for his
problem; it can be derived by filling in intermediate steps in Zhang's
calculation.}
\begin{eqnarray}
{d {\cal M}\over dt}  &=& {d W({\cal Q}, {\cal E})\over dt} + {d
E_{\rm int} \over dt} \nonumber \\
&=& -{1\over2} {\cal E}_{ij} {d\over dt} {\cal I}^{ij} -{1\over10} {d\over dt}
({\cal E}_{ij} {\cal I}^{ij}) \nonumber \\
&=& 3 {\cal E} {d {\cal Q} \over dt} + {3\over 5} {d ({\cal E}{\cal
Q})\over dt}\;. 
\label{eq:dM}
\end{eqnarray} 
The following argument tells us how much of this mass-energy change goes into
work and how much into interaction energy:  (i) 
The interaction energy $E_{\rm int}$ can depend
only on the instantaneous stellar deformation and tidal field, so $dE_{\rm
int}/dt$ must always be a perfect differential.  By contrast, 
the rate $dW/dt$ that
work is done need not be a perfect differential.
(ii) In a physical situation (not ours)
where the tidal field is changing while the stellar configuration is constant,
to first-order in the tidal perturbation no work is done on the star, $dW/dt =
0$.  These two facts are sufficient to imply that
\begin{equation}
{d W({\cal Q}, {\cal E})\over dt} 
= -{1\over2} {\cal E}_{ij} {d\over dt} {\cal I}^{ij} 
= 3 {\cal E} {d {\cal Q} \over dt}\;, 
\label{eq:WorkRate}
\end{equation}
\begin{equation}
{dE_{\rm int}\over dt} = -{1\over10} {d ({\cal E}_{ij} {\cal I}^{ij})\over 
dt} =  {3\over 5} {d ({\cal E}{\cal Q}) \over dt}\;.
\label{eq:EintRate}
\end{equation}
In our thought experiment, where the tidal field is fixed at $\cal E$ and the
quadrupole moment grows from zero to $\cal Q$, the total work done on the star
is 
\begin{equation}
W({\cal Q, \cal E}) 
= - {1\over2} {\cal I}_{ij} {\cal E}^{ij} = 3 {\cal Q}{\cal E}\;.
\label{eq:MQE}
\end{equation}

By inserting expressions (\ref{eq:MQE}) and (\ref{eq:MQ2}) into  
(\ref{eq:V}) and minimizing the resulting potential energy 
$V(N,R,{\cal Q,E})$ with
respect to $\cal Q$ at fixed $N,R,{\cal E}$, we deduce the equilibrium value of
the quadrupole moment: 
\begin{equation}
{\cal Q} = {3 R^5\over 2 \beta}{\cal E},
\label{eq:Qe}
\end{equation}
Correspondingly, the
potential energy of the quadrupolar equilibrated configuration is
\begin{equation}
V(N,R,{\cal E}) = M(N,R) - {9R^5\over4\beta}{\cal E}^2\;.
\label{eq:ME2}
\end{equation}

Because the quadrupolar contribution $- (9R^5/4\beta) {\cal E}^2$ to this
potential energy depends on the star's radius, the tidal coupling 
influences the star's radial, secular motions.  To deduce that influence, we
convert the potential energy to dimensionless units, 
\begin{equation}
v(n,r,\varepsilon) \equiv {V(N,R,{\cal E}) - M_o \over M_o}\;, \quad
\varepsilon \equiv {{\cal E}\over M_o/{R_o}^3}\;
\label{eq:eps}
\label{eq:veps}
\end{equation} 
and combine with Eqs.\ (\ref{eq:ME2}) and (\ref{eq:mnr}) to obtain
\begin{equation}
v(n,r,\varepsilon) = ar^3 + (b_0 + b_1 r)n - (c_0 + c_1 r)\varepsilon^2\;.
\label{eq:vnre}
\end{equation}
Here $\varepsilon$ 
is the external tidal field measured in units of the tidal field produced by
the critical configuration near its own surface, and 
\begin{eqnarray}
c_0 &=& {1\over M_o} \left( {M_o\over {R_o}^3} \right)^2 {9{R_o}^5 \over 4\beta}
= {9M_o\over 4\beta R_o} \sim 0.3\;, \nonumber \\ 
c_1 &=&  {R_o\over M_o} \left( {M_o\over {R_o}^3} \right)^2 
\left[ {d\over dR}\left( {9R^5\over 4\beta}\right)\right]_o 
\simeq 5 c_0 \sim 1.5\;,
\label{eq:c0c1}
\end{eqnarray}
where the factor 5 comes from differentiating the $R^5$.  The dependence 
of $\beta$ on the star's internal structure
may give rise to a slight variation of $\beta$ with $R$, which may slightly
change the relationship $c_1 = 5 c_0$; hence the approximate equality in Eq.\
(\ref{eq:c0c1}).

As the near critical star changes its radius $R$ slowly and secularly at fixed
tidal field $\cal E$, its
quadrupole moment continually equilibrates, with an accompanying tidal
feeding of energy $W$ into and out of the star.  
These radial motions conserve the star's relativistic kinetic energy plus its
potential energy $V(N,R,{\cal E})$ [which includes a correction for the
flow of $W$; cf.\ Eq.\ (\ref{eq:V})].  Correspondingly, the
star's equilibria are the extrema of $V(N,R,{\cal E})$---or equivalently
of its dimensionless version $v(n,r,{\varepsilon})$---and these equilibria are
stable if they minimize $v$ and unstable if they maximize it. 

By differentiating Eq.\ (\ref{eq:vnre}) with respect to $r$ at fixed $n$ and
$\varepsilon$, we obtain for the equilibrium configurations
\begin{eqnarray}
r_e &=& \pm \sqrt{(1/3a)(-b_1 n_e +c_1 \varepsilon^2)} \nonumber \\
&\sim& \pm 
{0.7 {\cal E} \over M_o/{R_o}^3}\hbox{ near criticality, where } n_e 
\ll {c_1\over b_1} \varepsilon^2\;.
\label{eq:re}
\end{eqnarray}
The configurations on the $+$ branch ($R>R_o$) are stable, and those on
the $-$ branch ($R<R_o$) are unstable.  

Notice that, in response to the tidal
field, each stable equilibrium configuration increases its radius ($r_e>0$); 
and correspondingly, since the fundamental radial-mode eigenfunction 
$\vec \xi ( \vec x)$ that describes the radial shape of its spherical 
deformation has
no nodes, its central density goes down.  Not surprisingly, this leads to
a secular stabilization of the star:  whereas before the tidal field was turned
on, the maximum number of baryons that the star could support without
collapsing was $N_o$, afterward the maximum baryon number has 
increased by a fractional amount
\begin{equation}
n_{e\; \rm max} = {c_1\over b_1}\varepsilon^2 
\sim 2.5 \left({{\cal E}\over M_o/{R_o}^3}
\right)^2\;.
\label{eq:nemax}
\end{equation}
 
The tidally induced increase of equilibrium radius 
(\ref{eq:re}) and increase of maximum baryon number
(\ref{eq:nemax}) are the same in sign and order of magnitude as Lai has 
deduced previously using post-Newtonian arguments \cite{Lai}.  

The inability of
fully relativistic gravity to produce a secular instability can be traced to 
the
robustly positive sign of the coefficient $c_1$, which in turn follows from the
fact that the star's deformation energy at fixed quadrupole moment, 
$\delta M({\cal Q},R)/{\cal Q}^2 = \beta/R^5$ is a decreasing function of
radius.  It is very hard to imagine any neutron star for which this would 
fail to be true. 

\section{Rotating Star and Rotating Tidal Field}
\label{sec:Rotation}

We now turn attention to a rigidly rotating neutron star interacting with 
a rotating external
tidal field, such as those which occur in binary neutron star systems.  
In our analysis, the rotation is with respect to the star's local asymptotic
rest frame, which itself will generally rotate with respect to ``infinity'' 
due to dragging of inertial frames by the binary's total angular momentum. 

For simplicity, we shall require that the star's spin axis and the tidal
field's rotation
axis coincide, as will be the case in a binary if the two
stars' spin axes are perpendicular to the orbital plane.  This requirement
protects the tidal field and the spin axis from precessing.  

The tidal stabilization analysis of Sec.\ \ref{sec:Static} can be generalized
to such a rotating system with little change.  We shall sketch that
generalization, beginning in Subsec.\ \ref{sec:RotatingStar} with the secular
stability of the rotating star in the absence of the tidal field, followed in
Subsec.\ \ref{sec:RotatingTidal} with the star-tide interaction and tidal
stabilization.

\subsection{Rotating Star Without Tidal Field}
\label{sec:RotatingStar}

Consider a family of rigidly rotating neutron stars that are characterized 
by the equation of state $P(\rho)$ and that all have the same spin angular
momentum $J$ as measured by frame dragging in their local asymptotic rest
frames.  For such stars, because $J$ is fixed once and for all,
the equilibrium configurations form a one-parameter family analogous to that
for static stars.  We shall characterize those equilibria by
their masses $M_e$ (as measured by Keplerian orbits in their local 
asymptotic rest frames), their number of baryons $N_e$, and their
equatorial radii $R_e$ (defined as their equatorial circumferences$/2\pi$). 

These equilibria will have a mass-radius relation $M_e(R_e)$ with the same
general shape as that for nonrotating stars (Fig.\ \ref{fig:MeRe}), but with a
critical mass $M_o$ and critical radius $R_o$ that are larger due to
centrifugal stabilization and centrifugal forces; cf., e.g., Figs.\ 1 and 2 of
Ref.\ \cite{HartleThorne}.

As for static stars, so for these spinning stars,
the critical configuration (configuration of maximum mass) is marginally 
stable to a secular expansion or
contraction that takes it horizontally in Fig.\ \ref{fig:MeRe} to another
equilibrium configuration with the same mass $M_o$ and angular momentum $J$;
the equilibrium configurations on the larger-radius side of critical 
are secularly stable,
while those on the smaller-radius side are secularly unstable.  

In these rotating stars, the motions of interest are {\sl secular} 
in two senses:
(i) As for static stars, the motions must be slow enough for pycnonuclear
reactions to go to completion so each element of expanding and contracting 
stellar material follows the same equation of state $P(\rho)$ as characterizes
the equilibrium stellar structure.  (ii) The motions must also be slow enough
for viscosity to produce enough coupling between adjacent mass elements
to keep the star rigidly rotating.  This second condition is especially 
severe; so
in practice the secular stabilization that we shall prove is of interest
primarily because of its implications for dynamical stability 
(Sec.\ \ref{sec:conclusion}). 

As for static stars, we construct from our one-parameter rigidly rotating 
family of equilibria a two-parameter family of configurations that are 
expanded and
contracted away from equilibrium.  Our construction procedure is precisely 
the same as in the
static case, with the slow, fundamental secular mode of deformation governing
the expansions and contractions.  The resulting deformed configurations will
obviously have mass-radius curves, $M(N,R)$ at fixed $N$ (and forever 
fixed $J$), with the same general
shapes as in the static case: the dashed curves of Fig.\ \ref{fig:mnr}. 
Moreover, when expressed in dimensionless variables 
(\ref{eq:dimensionless}), the $m(n,r)$ relation 
near the critical point for our rotating family
will have the same mathematical form, 
$m=ar^3 +(b_0+b_1r)n$, as for the static case.  The coefficients $a$,
$b_0$, and $b_1$ will be affected by the rotation and thus will not take 
on their static forms (\ref{eq:coefficients}), but in order to produce the 
equilibria's known stability properties, they will still all be positive.

\subsection{Tidally Deformed, Rotating Star}
\label{sec:RotatingTidal}

We now place each of our rigidly rotating, near-critical stars---all with 
the same
spin angular momentum $J$---in a rotating, external
tidal gravitational field.  We assume that the star and the tidal field 
rotate about the
same axis as seen in the star's local asymptotic rest frame; the star's
rotational angular velocity is $\omega \alt 2\pi\times 1000$/s, and if the 
rotating field
is produced by a binary companion, its angular velocity is that of the 
orbital motion, $\Omega \alt 2\pi \times 600$/s \cite{LaiWiseman}.  In this 
case, the tidal field is
exceedingly unlikely to excite the star's normal modes significantly.  This is
because (i) significant excitation requires resonant coupling; (ii) the 
circularly polarized $f$- and $p$-modes have angular velocities of pattern
rotation $\sigma \agt 2\pi \times 1500$/s (unless the star is rotating close to
centrifugal breakup), which is too large to resonate with the driving 
force except under the most extreme of circumstances; and 
(iii) the low-frequency $g$-modes, which {\sl can} resonate, 
have only very weak coupling to the tidal field 
\cite{Lai1,ReiseneggerGoldreich,Shibata}. 
With this justification, we shall assume that the tidal field does not
resonantly excite the star's normal modes.

The rotating tidal field will, however, raise a nonresonant, rotating 
quadrupolar tide on the neutron 
star.  If the star's material had zero viscosity, the tide would be 
perfectly aligned 
with the tidal field; i.e., the axes of the star's rotating quadrupole 
moment ${\cal I}_{jk}$ would be identical to those of the tidal field ${\cal
E}_{jk}$.  This standard Newtonian result must be true also in general
relativity since the relevant physics is nothing but that of simple harmonic
oscillators:  Whenever an undamped oscillator is driven by an off-resonance 
sinusoidal force, the oscillator's displacement response is precisely in
phase with the force.  As for an oscillator, so for the star, any small 
viscosity will cause a slight phase lag between excitation and response: the
orientation of the star's tidal deformation and its quadrupole moment will lag 
that of the tidal field by an angle 
\begin{equation}
\phi \simeq {(\Omega - \omega)\over (\sigma - \omega)^2 \tau_*}\;,
\label{eq:phi}
\end{equation}
where $\tau_*$ is the viscous damping time for the star's quadrupolar
excitations.  For realistic viscosities, this is an exceedingly small lag
angle \cite{Lai1,ReiseneggerGoldreich,Shibata}; 
so the quadrupole moment is very 
nearly oriented along the tidal field.  

This agreement of orientations prevents the tidal field from torquing the star,
so the star's spin angular momentum $J$ is conserved.

This conservation of $J$ and alignment of tidal field and quadrupole moment
enable us to carry over the static-star analysis of tidal stabilization
(Sec.\ \ref{sec:StaticTidal}) to the rotating case, essentially without change.
The tidal field and quadrupole moment can be characterized by scalars in the
manner of Eq.\ (\ref{eq:EQ}); raising the tide requires an energy $\delta
M(Q,R)$ given by Eq.\ (\ref{eq:MQ2}); the tide extracts from the tidal field an
amount of energy $W(Q,{\cal E})$ computable by integrating the Landau-Lifshitz
pseudotensor over a closed 2-surface in the star's local asymptotic rest frame
in the manner of Eq.\ (\ref{eq:LLint}), and that integration must give 
expression
(\ref{eq:MQE}) aside from a multiplicative factor that is close to 
unity.  This multiplicative factor 
(call it $1+\gamma$), which we have not computed,
will arise from the synchronous rotation of the tide and star with velocity
$\omega r$ at the star's surface.  It cannot differ from unity by more than
$\gamma \sim \omega^2 R^2 \alt 0.01$ (there is no linear term $\gamma \sim
\omega R$ because such a term would reverse sign when the star reverses
direction of rotation.)  As in the static case, 
the quadrupole moment adjusts itself so as to minimize the
potential energy function $V(N,R,{\cal Q}, {\cal E})$ of Eq.\ (\ref{eq:V}); 
the resulting $Q$ is expression (\ref{eq:Qe}) aside from the multiplicative
factor $1 + {\rm O}(\omega^2 r^2)$;
and the star's radius changes slowly and secularly in a manner governed
by the potential energy function $v(n,r,\varepsilon) = ar^3 + (b_0 + b_1 r)n -
(c_0 + c_1 r)\varepsilon^2 $ [Eq.\ (\ref{eq:vnre})].
The only difference is in the numerical values of the coefficients $a$, $b_0$,
$b_1$, $c_0$, and $c_1$.

As in the static case, the key to tidal stabilization is the sign of $c_1$; and
as there, so long as the star's deformation energy at fixed quadrupole moment,
$\delta M(Q,R)/Q^2 = (\beta/R^5) \times [1+ {\rm O}(\omega^2 R^2)]$ is a 
decreasing function of radius $R$, the
tidal field secularly stabilizes the star, increases its radius, and
decreases its central density.  It is exceedingly difficult to imagine a
star for which (when $\omega \alt 2\pi \times 600/$s 
as it must be in an inspiraling
binary \cite{LaiWiseman}) this would not be the case.

\section{Conclusions}
\label{sec:conclusion}

In this paper we have proved that tidal fields tend to stabilize a star against
secular gravitational collapse.  However, for spinning stars in binaries,
secular stability is irrelevant because radiation reaction drives the
binary through the regime of relativistic gravity far too
quickly for viscosity to keep the rotation rigid in pulsational motions.  
The relevant issue in this case is dynamical stability.

As is well known (e.g., Ref.\ \cite{FriedmanIpserSorkin}), secular stability
implies dynamical stability.  This is so for two reasons. First, 
pycnonuclear reactions go to completion in secular motions but not in 
dynamical motions, and as a result the secular motions are characterized by a
softer equation of state than the dynamical motions and thus are less stable. 
Second, dynamical motions of an initially rigidly rotating star produce
differential rotation; and the viscous coupling that converts that differential
rotation into the rigid rotation of a secular motion will necessarily extract
energy from the rotation.  This means that, beginning with the same
equilibrium configuration, the dynamical motion must lead to a configuration of
greater potential energy $V$ than the secular motion, which in turn means that
the equilibrium configuration is more stable against the dynamical motion than
against the secular motion.  

Consider a rigidly rotating neutron star in the final, relativistic phase of
binary inspiral.  Such a star must have lived for millions of years to 
reach this inspiral phase, so it must be secularly stable as well as
dynamically stable before the tidal field of its companion begins to affect it
significantly.  The growing tidal field, as we have seen, must increase the
star's secular stability.  Since secular stability implies
dynamical stability, the tidally deformed star must also remain dynamically
stable.  

This conclusion is not at all obvious from the equations of the
post-Newtonian approximation, as they are usually written: in a reference
frame that is asymptotically inertial at infinity.  One can identify in those
equations terms that appear able to produce tidal instability at first
post-Newtonian order \cite{WilsonMathews}.  This fact illustrates the
superficially
misleading character of the post-Newtonian equations:  Apparently magical
cancellations \cite{Wiseman} enforce the strong equivalence principle, 
which is fundamentally at the heart of tidal stabilization.  

By contrast, the
mathematical tools used in this paper, being based on the local asymptotic 
rest frame of the star whose stability interests us, are closely linked to the
strong equivalence principle and lead rather directly to the tidal
stabilization.  These tools are not widely used in relativistic astrophysics.
They are worth trying whenever one is interested in the behavior of 
a semi-isolated portion of a larger relativistic system---e.g., a neutron 
star or black hole interacting with other bodies \cite{ThorneHartle}.

As this paper was being completed, I learned of a similar,
local-asymptotic-rest-frame  analysis by Flanagan \cite{Flanagan}, which,
however, focuses on equations of motion rather than energy considerations.

\section*{Acknowledgments}

For helpful discussions and/or comments on the manuscript,
the author thanks Roger Blandford, Patrick
Brady, Doug Eardley, Eanna Flanagan, Scott Hughes, Dong Lai, Patricia Purdue,
Stuart Shapiro, 
Alan Wiseman, and an anonymous referee.  This research was supported 
in part by NSF Grant AST-9417371.

\section*{Appendix}

In this appendix, we elucidate the energetics of quadrupole-tidal coupling,
Eqs.\ (\ref{eq:LLint})--(\ref{eq:MQE}), by
working out its details in the Newtonian limit without rotation.  

We consider a nonroting Newtonian star that is spherical, aside from a
quadrupolar deformation and quadrupole moment ${\cal I}_{jk}$ 
that are aligned with an external tidal field ${\cal E}_{jk} = 
\Phi_{{\rm e},jk}$.  Here $\Phi_{\rm e}$ is the external Newtonian potential,
and ${\cal E}_{jk}$ and ${\cal I}_{jk}$ are symmetric and trace free.
We denote by $\rho$, $p$, $v$ and $\Pi$ the (Newtonian) mass density,
pressure, velocity and specific internal energy of the stellar fluid,
and by 
\begin{equation}
\Phi = \Phi_{\rm o} + \Phi_{\rm e}
\label{eq:PhiSplit}
\end{equation}
the Newtonian gravitational
field and its split into the star's self field $\Phi_{\rm o}$ and the external
field $\Phi_{\rm e}$. 

In Newtonian theory, the total energy density and energy flux of the stellar
fluid plus gravitational field can be written as
\begin{eqnarray}
\Theta^{00} &=& \rho \left( {1\over2}v^2 + \Pi + \Phi\right) + {1\over8\pi} 
\Phi_{,j} \Phi_{,j}\;, \label{eq:theta00} \\
\Theta^{0j} &=& \rho v^j \left( {1\over2}v^2 + \Pi + {p\over\rho} + \Phi\right)
- {1\over4\pi} \Phi_{,t}\Phi_{,j}\;,
\label{eq:theta0j}
\end{eqnarray}
where we make no distinction between covariant and contravariant spatial indices
because our coordinates are assumed Cartesian.  It is straightforward to verify
that the energy conservation law
\begin{equation}
{\Theta^{00}}_{,t} + {\Theta^{0j}}_{,j} = 0
\label{eq:divtheta}
\end{equation}
is satisfied by virtue of mass conservation $\rho_{,t} + (\rho v^j)_{,j} = 0$,
the first law of thermodynamics, $\rho d\Pi/dt + {p v^j}_{,j} = 0$, the fluid's
Euler equation $\rho dv^j/dt + \rho \Phi_{,j} + p_{,j} = 0$, and Newton's
field equation $\Phi_{,jj} = 4\pi\rho$.  Here $d/dt = \partial/\partial t +
v^j \partial/\partial x^j$ is the fluid's comoving time derivative.  

[{\it Note:} 
Equations (\ref{eq:theta00}) and (\ref{eq:theta0j}) entail a specific choice
of how to localize the system's gravitational energy.  Other choices are
possible: one can add to $\Theta^{00}$ the divergence of $\eta_j \equiv \alpha
\Phi \Phi_{,j}$ 
(where $\alpha$ is an arbitrary constant) 
and add to $\Theta^{0j}$ the time derivative of $\eta_j$ 
without affecting the law of energy
conservation (\ref{eq:divtheta})
or any of the system's physics.  This nonuniqueness of 
localization of gravitational energy also occurs in general relativity; cf.\
Chap.\ 20 of MTW\cite{MTW}.  We shall discuss its consequences below.]

The external field is purely quadrupolar and source free throughout the star
and the star's local asymptotic rest frame
\begin{equation}
\Phi_{\rm e} = {1\over2} {\cal E}_{ij} x^i x^j\;, \quad \Phi_{{\rm e},jj} =
0,
\label{eq:Phie}
\end{equation}
and its tidal field ${\cal E}_{ij}$ evolves with time.  The star's external self
field is purely monopolar and quadrupolar and its source, of course, is the
star's mass distribution
\begin{equation}
\Phi_{\rm o} = - {M\over r} -{3\over2} {{\cal I}_{ij}n^i n^j \over r^3}
\hbox{\rm outside star,} \quad \Phi_{{\rm o},jj} = 4\pi\rho\;.
\label{eq:Phio}
\end{equation}
Here $r \equiv \sqrt{\delta_{ij} x^i x^j}$ is radius and $n^j \equiv x^j/r$ is
the unit radial vector.  The star's mass $M$ is constant in time, but its
quadrupole moment ${\cal I}_{ij}$ evolves.

Now consider the total energy $E$ inside a ball $\cal V$ centered on the star
and larger than the star.  This energy consists of three parts, the star's 
self energy $E_{\rm o}$, the external field's energy $E_{\rm e}$, and 
the energy $E_{\rm int}$ of interaction between the
star's quadrupolar deformation and the external tidal field
\begin{equation}
E \equiv \int_{\cal V} \Theta^{00} d^3x = E_{\rm o} + E_{\rm e} + 
E_{\rm int}\;,
\label{eq:E}
\end{equation}
\begin{eqnarray}
E_{\rm o} &=& \int_{\cal V}\left[ 
\rho\left( {1\over 2} v^2 + \Pi + \Phi_{\rm o} \right) 
+ {1\over 8\pi} \Phi_{{\rm o},j}\Phi_{{\rm o},j} \right] d^3x\;, 
\label{eq:Eo} \\ 
E_{\rm e} &=& \int_{\cal V} {1\over 8\pi} \Phi_{{\rm e},j} \Phi_{{\rm e},j} 
d^3x\;, 
\label{eq:Ee} \\
E_{\rm int} &=& \int_{\cal V} \left( \rho \Phi_{\rm e} 
+ {1\over 4\pi} \Phi_{{\rm o},j} \Phi_{{\rm e},j} \right) d^3x 
= {3\over 10} {\cal E}_{ij} {\cal I}_{ij}\;.
\label{eq:Eint} 
\end{eqnarray}
Here the value of $E_{\rm int}$ follows from the form and sourcelessness
(\ref{eq:Phie}) of the
external field, the source equation and exterior form of the star's self field
(\ref{eq:Phio}), and the usual volume integral for the quadrupole moment
${\cal I}_{ij} = \int_{\cal V} \rho ( x^i x^j - {1\over 3} r^2 \delta^{ij} )
d^3x$.

The law of local energy conservation (\ref{eq:divtheta}) guarantees that the
rate of change of the total energy (\ref{eq:E})
is the surface integral of the energy flux
over the boundary $\partial {\cal V}$ of the ball $\cal V$  
\begin{equation}
{dE\over dt} = - \int_{\partial{\cal V}} \Theta^{0j} n^j r^2 d\Omega\;. 
\label{eq:dEdtsurf}
\end{equation}
Here $d\Omega$ is solid angle.
By combining Eqs.\ (\ref{eq:dEdtsurf}),
(\ref{eq:theta0j}), (\ref{eq:PhiSplit}), (\ref{eq:Phie}),
(\ref{eq:Phio}) and (\ref{eq:Ee}),
we find that
\begin{eqnarray}
{dE\over dt} &=& {dE_{\rm e}\over dt} 
+ \int_{\partial {\cal V}} 
{1\over 4\pi} \Phi_{{\rm o},t} \Phi_{{\rm o},j} n^j r^2 d\Omega
\nonumber \\
&&+ {d\over dt}\left({3\over10} {\cal I}_{jk} {\cal E}_{jk} \right) - {1\over 2}
{\cal E}_{jk} {d\over dt}{\cal I}_{jk} \;.
\label{eq:dEdt}
\end{eqnarray} 
The first term is the rate of change of the external field energy, inside the
ball, due to the evolution of the tidal field ${\cal E}_{jk}$.  The second term
is the rate of change of the star's self energy due 
its own field energy flowing into or out of the ball $\cal V$ as its quadrupole
moment decreases or increases; 
in the limit as the 
ball's radius becomes arbitrarily large, this term goes to zero.  The third
term is the rate of change of the interaction energy; cf.\ Eq.\ (\ref{eq:Eint}).
By comparing with Eq.\ (\ref{eq:E}), we see that the last
term must drive changes in the star's self energy
\begin{equation}
{dE_{\rm o}\over dt} = \int_{\partial {\cal V}} 
{1\over 4\pi} \Phi_{{\rm o},t} \Phi_{{\rm
o},j} n^j r^2 d\Omega - {1\over 2}
{\cal E}_{ij} {d\over dt}{\cal I}_{ij}\;;
\label{eq:dEodt}
\end{equation}
In other words, the last term is the rate at which the external tidal field 
does work on the star
\begin{equation}
{d W({\cal Q}, {\cal E})\over dt}
= -{1\over2} {\cal E}_{ij} {d\over dt} {\cal I}^{ij}\;;
\label{eq:WorkRate1}
\end{equation}

This work rate is in perfect agreement with the fully relativistic work rate
(\ref{eq:WorkRate})
derived in the text using the Landau-Lifshitz pseudotensor to
localize the gravitational energy and using deDonder 
(harmonic) gauge. 

Not so the rate of change of interaction energy.  The Newtonian analysis gives
\begin{equation}
\left({ dE_{\rm int}\over dt}\right)_{\rm Newton} = 
{3\over10} {d ({\cal E}_{ij} {\cal I}^{ij})\over
dt}   
\label{eq:EintRateNewton}
\end{equation}
[Eq.\ (\ref{eq:Eint}) above]; by contrast, the relativistic
analysis gives this same expression but with a coefficient $-1/10$ rather than
$+3/10$; see Eq.\ (\ref{eq:EintRate}).  This disagreement does not arise from
any fundamental difference between Newtonian theory and general relativity. 
Rather, it arises because the localization of gravitational energy is
non-unique, both in Newtonian theory and in general relativity [see the
paragraph preceding Eq.\ (\ref{eq:Phie})]; 
and  $E_{\rm int}$ is that portion of the system's total 
gravitational interaction 
energy which resides in the vicinity of the particular 
neutron star on which we are focusing. For a binary system,
our two analyses distribute the
gravitational energy differently between the stars' vicinities and the  
interstellar region, and they thereby give rise to different values of $E_{\rm
int}$.  If, in the Newtonian analysis, we were to change our localization of 
the gravitational energy via the transformation described in the paragraph
preceding Eq.\ (\ref{eq:Phie}), 
we would alter $E_{\rm int}$, while leaving unchanged the uniquely
defined work $W$ done by the tidal field on the star of interest.  Similarly,
if, in our general relativitistic analysis, we were to change our energy
localization by switching from the Landau-Lifshitz pseudotensor to some other
pseudotensor, or by performing a gauge change on the gravitational field,
we thereby would alter $E_{\rm int}$ but leave $W$ unchanged.

This localization dependence of $E_{\rm int}$ and uniqueness of $W$ 
will be analyzed in detail in a subsequent paper \cite{PurdueThorne}.


\end{document}